\documentclass[runningheads]{llncs}
\usepackage{amsmath,amssymb}
\usepackage{graphicx}
\usepackage{multirow}
\usepackage{enumerate}
\usepackage{url}
\begin{document}
\title{Malicious Web Request Detection Using Character-level CNN}
%
%
\author{Wei Rong\inst{1} \and
	Bowen Zhang\inst{1} \and
	Xixiang Lv\inst{1}}
\authorrunning{Wei Rong et al.}
%
\institute{Xidian University, 266 Xinglong Section of Xifeng Road, Xi’an, Shaanxi 710126 , China}
\maketitle              
\begin{abstract}
Web parameter injection attacks are common and powerful. In this kind of attacks, malicious attackers can employ HTTP requests to implement attacks against servers by injecting some malicious codes into the parameters of the HTTP requests. Against the web parameter injection attacks, most of the existing Web Intrusion Detection Systems (WIDS) cannot find unknown new attacks and have a high false positive rate (FPR), since they lack the ability of re-learning and rarely pay attention to the intrinsic relationship between the characters. In this paper, we propose a malicious requests detection system with re-learning ability based on an improved convolution neural network (CNN) model. We add a character-level embedding layer before the convolution layer, which makes our model able to learn the intrinsic relationship between the characters of the query string. Further, we modify the filters of CNN and the modified filters can extract the fine-grained features of the query string. The test results demonstrate that our model has lower FPR compared with support vector machine (SVM) and random forest (RF).

\keywords{Malicious Detection \and Injection Attacks \and CNN \and Embedding \and Deep Learning.}
\end{abstract}

\section{Introduction}
\begin{figure}
	\includegraphics[width=10cm]{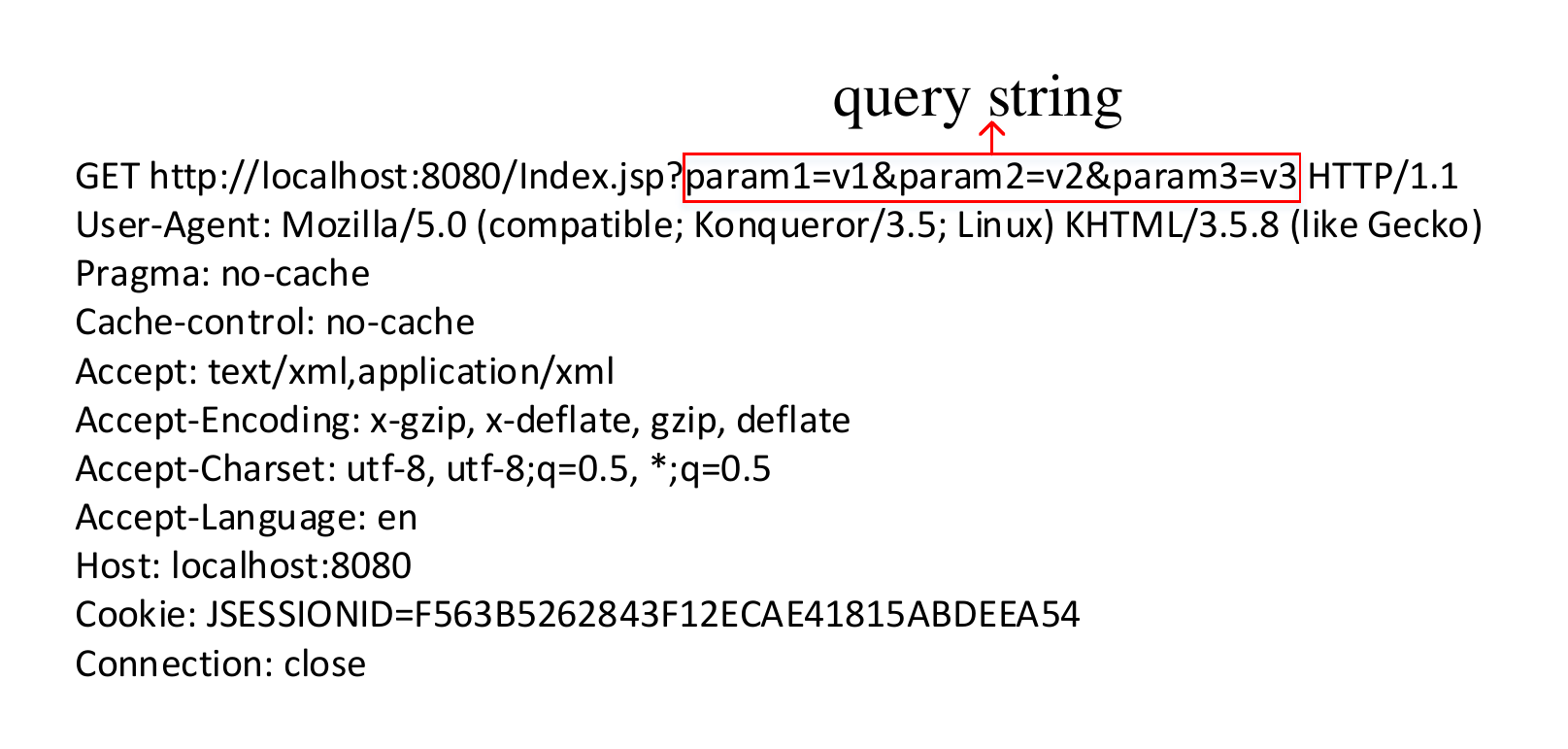}
	\centering
	\caption{a request parameter in an HTTP header} \label{Fig1}
\end{figure}

By HTTP/HTTPS, a user client can initiate a web request and send it to the web server. A web request usually carries some parameters input by users. Fig.~\ref{Fig1} shows the query string in GET method which is one of HTTP methods. Many attackers use this part to pass their malicious code to the web server.

Web application attacks are common in attack incidents against web servers. According to the recent report of Alert Logic~\cite{news,report2}, 73\% of the attack incidents in the past 18 months are web application attacks and these attacks affected 85\% of its customers. Akamai(a Content Delivery Network service provider) also published that hackers had launched 47,417,828 web attacks aiming at Akamai servers in just 8 days, between November 8, 2017 to November 15, 2017~\cite{report}. Within all web attacks, code injection attacks account for a large proportion. By injecting some malicious codes into the parameters of the HTTP requests, malicious attackers can employ HTTP requests to accomplish their attacks against servers. This type of attacks is called as parameter injection attacks. Thus, it is necessary for service providers to build an efficient mechanism to detect code injection attacks.

Against web parameter injection attacks, the existing method is Web Intrusion Detection Systems(WIDS), which includes signature-based detection and anomaly-based detection~\cite{wiki:Detection_method}. With signature-based detection, the server needs to maintain a library of malicious symbols that distinguish an injection attack from normal web requests. When it receives a request, the server searches the query string to see if they contain a malicious symbol, and a web request containing a malicious symbol is considered as an attack. This signature-based detection usually performs very well on the known attacks that contain distinct symbols. However, a tiny change on the injected codes will hide the symbols and disable this method. Obviously, signature-based detection cannot find unknown new attacks. As for anomaly-based detection, the server needs to train a math model which can characterize normal web requests and filter most abnormal web requests. Compared with signature-based detection, this type of WIDS is able to discover some unknown new attacks. Unfortunately, the FPR (false positive rate) of anomaly-based detection is high in practice.

Besides the high FPR, the model of anomaly-based detection lacks the ability of re-learning since it is a constant model. In fact, earlier anomaly-based detection system collects all training data once and then generates a constant model. Earlier anomaly-based detection systems take the sample with abnormal data structures as an attack. Thus, it can detect some unknown attacks. However, when the data structure of an attack is similar with a normal one, an anomaly-based detection system is incapable of finding the attack. The reason is that an anomaly-based WIDS cannot update its model in time.

In this paper, by using CNN we build a model for detecting parameter injection attacks in terms of web applications. In our model, we add a character-embedding layer before the convolutional layers. The parameters within a request is first fed into the character-embedding layer, and the character-embedding layer transforms them into feature vectors which are then put into a traditional CNN along with their labels. The output of CNN is a prediction demonstrating if the request is a malicious attack or not. The point is that our system is dynamic and has the ability of re-learning. That is, the model can be restored when there is some new data. Our system not only is able to find unknown new attacks, but also is able to extract features of the malicious query automatically instead of maintaining a feature library like the signature-based detection. Since our model can extract both normal features and malicious features, it has lower FPR than anomaly-based detection model which only learns normal features.

The main contributions of this paper are as follows:

We build an improved CNN model in order to detect malicious web requests. In this model, we add a character-level embedding layer before the convolution layer aiming to digitize a character sequence. Different from the normal digitization, character-level embedding is able to learn the hidden character similarity in web request. In the convolution layer, considering the one-dimensional data structure, we design vertical filters to extract features from character sequence.

Based on the improved CNN model, we present an online learning system  which has re-learning ability. That is, the model can be updated in time when there are some unknown new attacks. Therefore, it can adapt to the changing  attack methods.

\section{Related work}
\label{RELATED WORK}

\subsection{Malicious request detection}
Malicious request detection has been an active research area, and various malicious request detection systems have been proposed. They provide solutions for specific type or multiple types of malicious request. For example, reference~\cite{ml_sqli2,svm_sqli,ml_sqli} show us their solution to SQLI attack that is one of the most common malicious requests. Reference~\cite{svm_sqli,ml_sqli} train a Support Vector Machine(SVM) model to detect SQLI attacks. Joshi and Geetha~\cite{ml_sqli2} detect SQLI attacks using the Naive Bayes machine learning algorithm. Reference~\cite{ml_sqli2,svm_sqli} create feature vector of string by the tokens of sql-query strings and feeds these vectors into the SVM-Train process to generate a model. Uwagbole and Buchanan~\cite{ml_sqli} create feature vectors by a hashing procedure called Feature hashing. But these method of creating feature vectors cannot reflect the relationship between words. On the other hand, the model of these paper is a constant model that is hard to be updated when there are some new form of SQLI attacks. Dong and Zhang~\cite{dong2017adaptively} proposed an adaptive system based on SVM for detecting malicious queries in web attacks. This system focused on 4 most frequent types of web-based code-injection attacks. The model of this system can be updated by some new queries periodically. But it does not pay attention to the local features which have an important impact on the final detection result.

\subsection{Convolutional neural network}
In recent years, neural networks(NNs) have been successfully applied in many areas. To some extent, malicious web request detection is a special kind of text classification problem. NNs have achieved great success in text classification area. For example, Kim~\cite{kim2014convolutional} applied CNNs to sentence classification and Shibahara et al.~\cite{shibahara2017malicious} applied CNNs to url sequence which is similar to parameter string focused in this paper. Kim~\cite{kim2014convolutional} proposed a special CNNs named TextCNN. TextCNN converts word sequence to a matrix which CNNs are able to process through embedding layer. Shibahara et al.~\cite{shibahara2017malicious} proposed an Event De-noising CNN(EDCNN). The input of EDCNN is more than just url sequence. They use one-hot representation to denote the categorical features which contain city corresponding to the IP address, TLD and country corresponding to the IP address. In conclusion, CNN performs very well in these text classification problem, therefore it is possible to solve the malicious web request detection problem using CNN.

\section{Proposed system}
\begin{figure}
	
	\includegraphics[width=\textwidth]{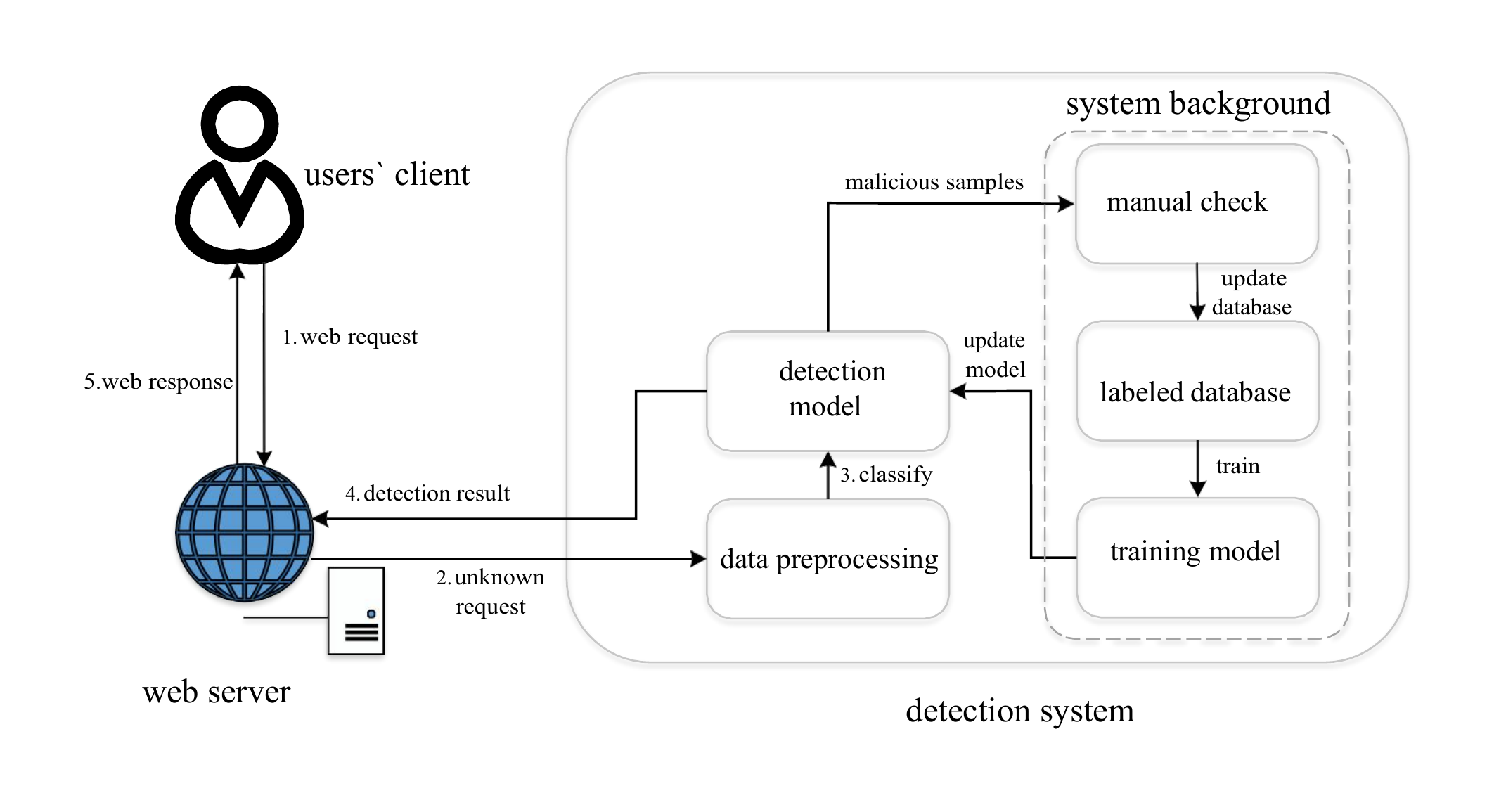}
	
	\caption{System Overview}
	\label{Fig2}
\end{figure}

In this section, we will present the proposed system detecting malicious web request and give its details.

\subsection{System design}

Our system receives the parameters of the web request as input and outputs the detection results. As shown in Fig.~\ref{Fig2}, the system includes five modules which are data preprocessing, classification model, manual check, labeled database and training model.

Before we run our system, we need some labeled samples and use them to train an initial detection model. As illustrated in Fig.~\ref{Fig2}, the system works as follows:

\begin{enumerate}[1)]
	\item A user client sends a web request with some parameters to the web server.
	\item The web server passes the request to the data preprocessing  module. The data preprocessing  module extracts the request parameter from the request and converts it into a index list by indexing these characters.
	\item The data preprocessing module feeds this index list into the detection model.
	\item After some sophisticated computations with respect to this vector, the detection model returns the detection result to the web server.
	\item The web server responds to the user client according to the detection result.  If it is a benign request, the web server will return a web response that it wants. Otherwise, the web server will return an error.
\end{enumerate}

In the system background, the classification model saves some samples and their detection results. These saved samples are randomly selected from those samples that the detection model is less confident about (has relatively low probability). In order to ensure that the labeled data is representative and the amount of each type of data is balanced, we should check the samples manually before adding them into the labeled database. When the system finds that the labeled database has changed, it will train a new detection model based on the old model by using the changed database. Then, system updates its detection model with the new model. By this way, the model can get the re-learning ability and adapt to the change of web attacks. 

In this system, the detection task is mainly completed by the detection model. The details of the detection model will be given in the next subsection.
\subsection{Detection Model}

Aiming to analyze web query string, we build an improved CNN model. Using character embedding, we make the model learn by itself to digitalize characters according to its tasks. Then, the improved CNN extracts the features of query string and gives the detection results on the output layer.
\subsubsection{Character representation}

\begin{figure}
	
	\includegraphics[width=\textwidth]{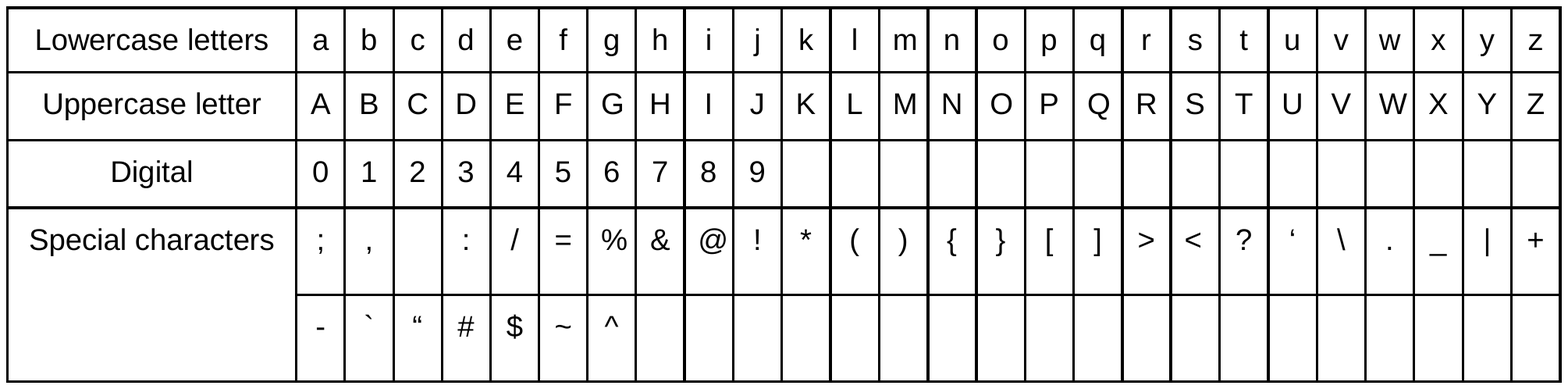}
	
	\caption{Characters List}
	\label{Fig3}
\end{figure}

\begin{figure}
	
	\includegraphics[width=10cm]{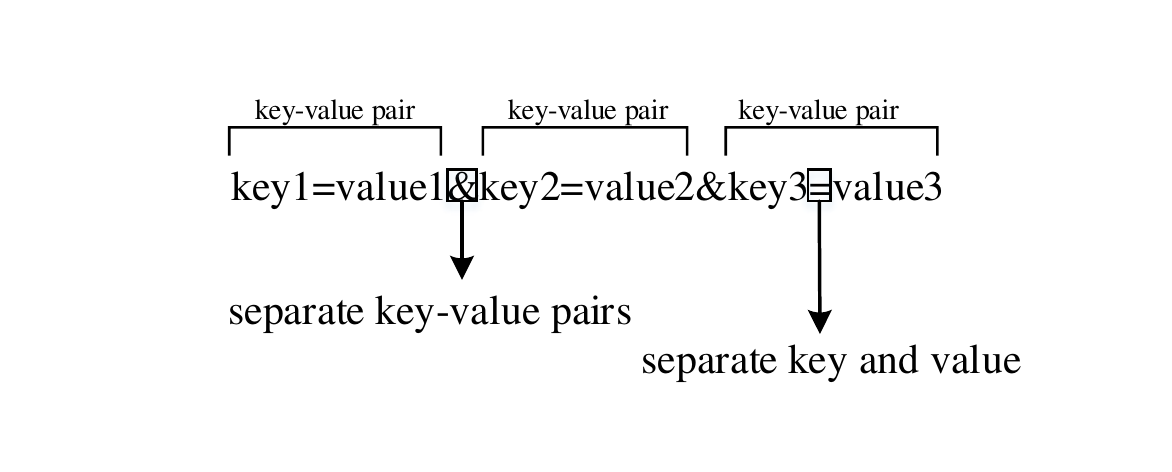}
	\centering
	\caption{the character `=' and `\&' in a request parameter}
	
	\label{Fig4}
\end{figure}

A request parameter is a character sequence which contains three types of characters, letters, digitals and special characters, which are illustrated in Fig.~\ref{Fig3}. There are 95 characters totally. We convert character sequences into index lists by indexing these characters. However, the character index cannot reflect the relationship between the characters in a specific context. For example, the character `=' is not associated with the character `\&' in normal texts. As shown in Fig.~\ref{Fig4}, in the context of web query string, they play the similar role which is a separater. This similarity connot be represented by index numbers.

In terms of this problem,we use character-level embedding in our model to learn the hidden similarity between characters in web query string. The idea of character-level embedding comes from the word embedding used in NLP~\cite{bengio2003neural}.

\subsubsection{Model Architecture}

\begin{figure*}[t]
	\centering
	\includegraphics[width=\textwidth]{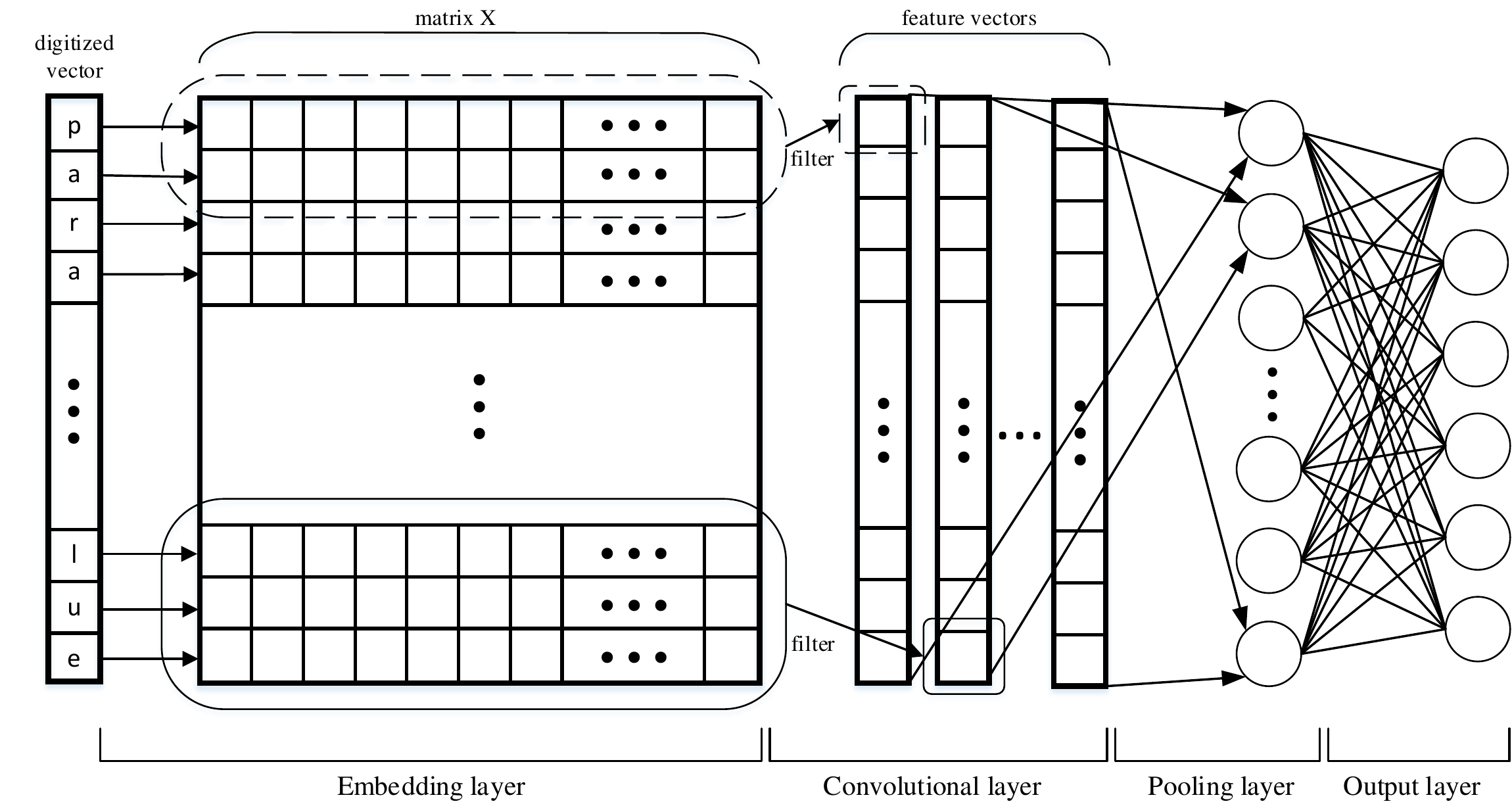}
	\caption{Model architecture}
	\label{Fig6}
\end{figure*}

\begin{figure}
	
	\includegraphics[width=8cm,clip]{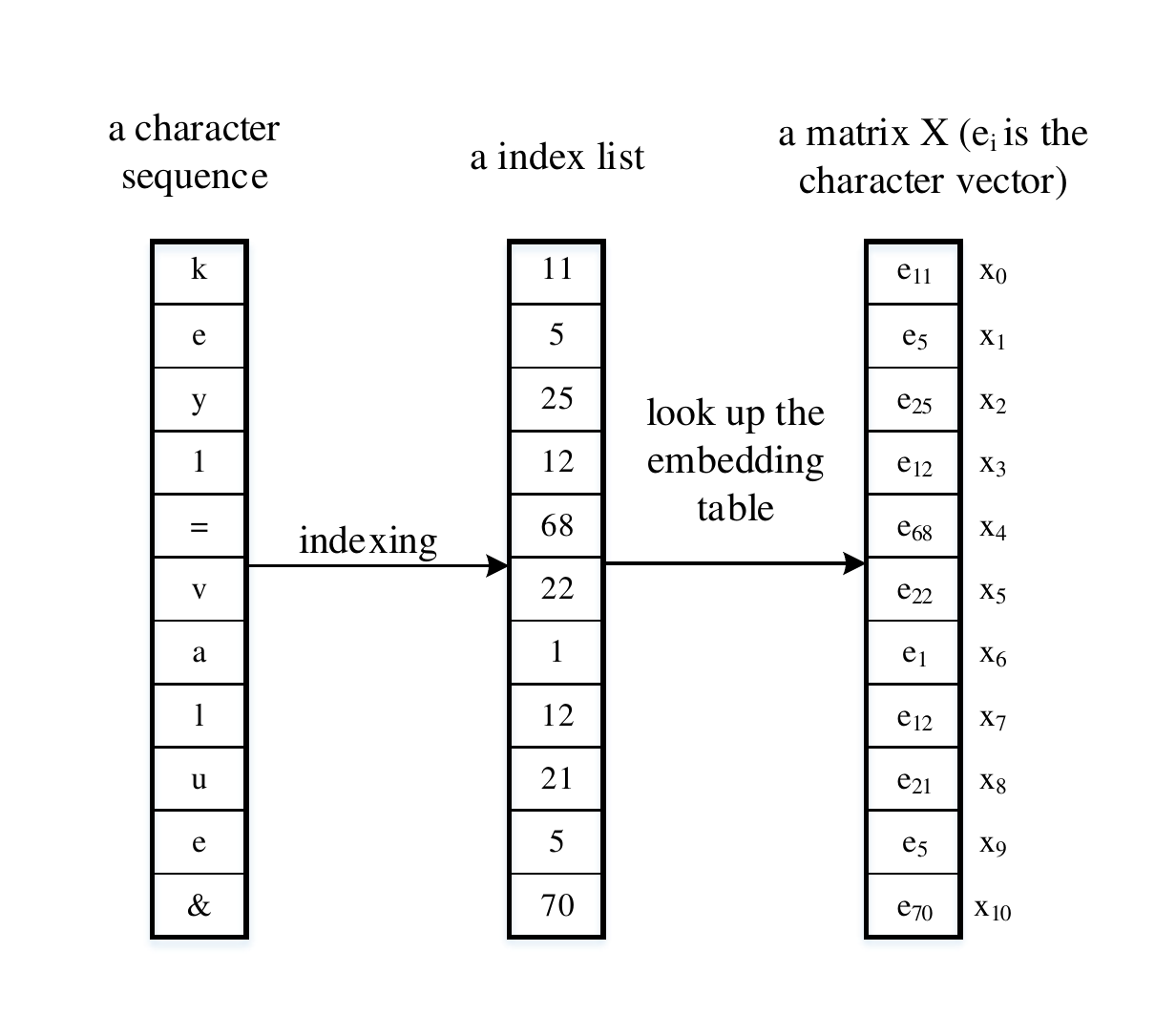}
	\centering
	\caption{convert a character sequence to a matrix X}
	\label{Fig5}
	
\end{figure}
The model architecture, shown in Fig.~\ref{Fig6}, is a deep neural network we used in our demo system. It mainly consists of four parts: the embedding layer, the  convolution layer, the pooling layer and the output layer. 

\textsl{Embedding layer:} The embedding layer converts a index list into digital matrix and then feeds it into the convolution layer. Let $e_i \in \mathbb{R}^k$  be the k-dimensional vector corresponding to the $i$-th character in the character list. Here, $i$ denotes the index of a character. Then, we get an embedding table $E = \begin{pmatrix} e_0, e_1, e_2,..., e_{94} \end{pmatrix}^T$, from which we can query every character vector by the character index. The embedding table is randomly initialized before starting training and will be optimized by backpropagation to minimize the classification error. As shown in Fig.~\ref{Fig5}, for a character in the character sequence, we look up in the embedding table $E$ and get a vector $e_i$ using the index of the character, thus making a matrix $X = \begin{pmatrix} x_0, x_1, x_2,..., x_{n-1} \end{pmatrix}^T$ for the whole character sequence, where $x_i^T \in E$. At the beginning of training, $x_i$ is randomly generated. After several iterations, $x_i$ will be replaced by a new $x_i^*$ which has been optimized by the backpropagation algorithm. If there are some internal relationships between the character $c_a$ and the character $c_b$ in the request parameter, the Euclidean distance between $x_a^*$ and $x_b^*$ will be smaller relatively. For example, in the experiment we found the Euclidean distance between `=' and `\&' ($L_2(x_4^*, x_{10}^*) = 7.8$) is smaller than the distance between `a' and `\&' ($L_2(x_6^*, x_{10}^*) = 10.09$), and in practice `=' and `\&' play the similar role.

\textsl{Convolution layer:} The convolution layer uses different filters to extract different feature vectors. The filters are some matrices with different sizes. But these matrices must have $k$ columns, where $k$ is the column of $E$ in Embedding layer. 
It should be noted that the difference in size only lays in rows. Unlike CNN applied in image recognition, the filters in our model can only be moved vertically. Because we want our CNN model to learn the relationship between characters, and every row of the matrix $X$ represents one character. In our model, we set the width of filters equal to that of matrix $X$ to ensure the indivisibility of each character. Let $y$ be one of the output feature vectors and $F$ be a filter. Then the output feature vector $y$ is computed as:
$$y_i = Relu(\sum_{r=0}^{L-1} \sum_{c=0}^{k-1} F_{r, c}X_{i+r, c})$$
where $y_i$ denotes $i$-th value and $F$ has $L$ rows and $k$ columns. Relu(Rectified Linear Units) function is a popular activation function in the field of deep learning, which is proposed and applied in reference~\cite{nair2010rectified}. When we use $n$ different filters $F$, we will get $n$ feature vectors $y$ in convolution layer and these vectors may have different lengths.

\textsl{Pooling layer:} Pooling is a commonly used method in the field of deep learning. Pooling layer outputs $n$ results, where $n$ denotes the amount of filters in convolution layer. In our system, we use max pooling to calculate the outputs. The output is derived as $z_i = max_{0\leqslant t \leqslant l-1}(y^i_t)$, where $l$ denotes the length of feature vector $y$ in convolution layer. This pooling layer is used to reduce the amount of parameters and the computation in our network and hence to control overfitting.

\textsl{Output layer:} The output layer outputs the classification results. The length of the output vector is the number of the classes. The output vector is computed as $y_j = Softmax(\sum_{i=0}^{n-1}w_{i, j}z_i)$, where $n$ denotes the length of the output vector of previous layer, $w$ denotes weights of this layer and the function Softmax is computed as $Softmax(x_j) = \dfrac{e^{x_j}}{\sum_i e^{x_i}}$. In order to improve the generalization ability of our model, we can also add dropout regularization in this layer. If we use dropout in this layer, $z_i$ we get from the previous layer has a 50\% probability of being zero. The summation of all output values of this layer is 1 and the output values are all non-negative. Therefore, an output value represents the probability of belonging to a specific class.

To sum up, the embedding table, the filter matrices and the weights of every layer are what we need to train, and they will be optimized by backpropagation algorithm to minimize classification error. 

\section{Test and Evaluation}

\subsection{Dataset}
\begin{table}  
	\caption{Composition of dataset}  
	\centering
	\begin{tabular}{|l|c|r|}
		\hline
		Category&Number of samples& Proportion\\
		\hline
		Benign request& 2000 & 46.49\% \\
		\hline
		SQLi& 472 & 10.97\% \\
		\hline
		XSS& 720 & 16.74\% \\
		\hline
		RFI& 599 & 13.92\% \\
		\hline
		DT& 511 & 11.88\% \\
		\hline
		SUM & 4302 & 100\% \\
		\hline
	\end{tabular}
\end{table}
In our experiment, we build our dataset through crawling from internet or extracting the data we need from other datasets~\cite{dataset2,CSIC}.  Then we merge them into one dataset. There are 5 types of data samples we chose in our dataset: Benign request, SQL injection(SQLI), Remote File Inclusion(RFI), Cross-Site Scripting(XSS) and Directory Traversal(DT). However, the amount of each category is extremely unbalanced. In order to keep the relative balance between each category, we set a threshold for the amount of each category. When the number of samples in one category is greater than this threshold, we will randomly sample this category to limit the number of samples to this threshold. Table 1 displays the composition of the dataset we use finally.
\subsection{Test Environment}
We implement our architecture in Python 3.5.3 using Tensorflow~\cite{tensorflow}. The operating system we use is Ubuntu 17.04. We also use some of Python's third-party libraries, scikit-learn~\cite{scikit-learn} and numpy.

\begin{table*}	  
	\caption{Performance of each model} 
	\centering
	\begin{tabular}{|c|c|c|c|c|c|c|c|}
		\hline
		Model &Indicators& Benign request &DT &RFI &SQLi &XSS & FPR\\
		\hline
		\multirow{2}*{RF}& recall &99.62\% & 99.99\%& 99.58\%&99.99\% &99.65\% &\multirow{2}*{0.38\%}\\
		\cline{2-7}
		~& precision & 99.99\%& 98.58\%& 99.57\%& 99.53\%&100\% &~\\
		\hline
		\multirow{2}*{SVM}& recall & 99.87\%& 99.53\%& 98.25\%& 98.52\%&99.99\%  &\multirow{2}*{0.13\%}\\
		\cline{2-7}
		~& precision & 99.12\%& 99.53\%& 99.98\%&99.53\% &100\% &~\\
		\hline
		\multirow{2}*{CNN}& recall & 99.95\%& 100\%& 100\%&100\% &100\% &\multirow{2}*{0.02\%}\\
		\cline{2-7}
		~& precision & 100\%& 100\%& 99.83\%\%&99.79\%&100\% &~\\
		\hline
	\end{tabular}
\end{table*}

\subsection{Evaluation}
\begin{figure}[bhtp]
	\centering
	\includegraphics[width=10cm]{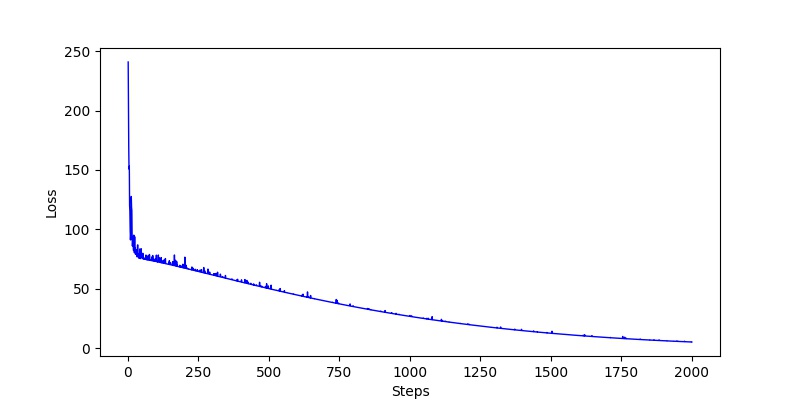}
	\caption{Train process}
	\label{Fig7}
\end{figure}

We implement a demo classifier for malicious requests detection using the model shown in Fig.~\ref{Fig6}. In the training process, we shuffle all training samples and divide them into several batches. Then the samples of each batch are used to optimize embedding table and weights of every layer. The process that we use one batch of samples to optimize these parameters is one step. So we can optimize these parameters by all batches step by step. Fig.~\ref{Fig7} shows how our model is optimized step by step and demonstrates the sum of softmax cross entropy and L2 regularization~\cite{ng2004feature}.

In order to compare with the traditional model, we add two control group. We chose Random Forest(RF) and SVM as the control group. The model of control group converts character sequences into sparse matrix through TF-IDF algorithm. We calculate precision and recall rate~\cite{wiki:Precision_and_recall} for every category and we also calculate FPR (we choose malicious samples as positive samples) for each model. So for one model we have one FPR and 5 pairs of precision and recall rate to calculate.  The results of our model against other two models are listed in Table 2. Our model distinctively reduces FPR from 0.38\% in RF and 0.13\% in SVM to 0.02\%. We attribute this to the character-level embedding layer and the modified filters of CNN. It should be noted that the limited improvement in the recall and precision rate of our model is due to the limitations of the dataset. Therefore, there is reason to believe that better results will be achieved when comes to practical issues. Because only one convolution layer is used in our demo system and more advanced features will be easily extracted if we use more convolution layers.

\section{Conclusion}
We proposed a malicious requests detection system with re-learning ability based on an improved CNN model. Compared with two other traditional model: SVM and RF, our model pays more attention to the extraction of the local feature in the query string. Specifically, we add a character-level embedding layer and modify the filters of CNN to extract the local feature in the query string. The results show that our improved CNN model outperforms traditional models on test datasets, while the performance in more practical applications remains to be tested.

\bibliographystyle{splncs04}
\bibliography{ref}
\end{document}